\documentclass[journal]{IEEEtran}

\usepackage{amsmath}
\usepackage{mathtools}
\usepackage{amssymb}
\usepackage{physics}
\usepackage{graphicx}
\usepackage{subfigure}
\usepackage{dblfloatfix}
\usepackage{cite}
\usepackage{textcomp}
\usepackage{makecell}
\usepackage{cases}

\begin{document}
\title{Two-Port Feedback Analysis On Miller-Compensated Amplifiers}

\author{Myungjun~Kim,~\IEEEmembership{Member,~IEEE}
\thanks{Myungjun Kim is with Samsung Electronics, Hwaseong,
Korea e-mail: phaedrus.kim@samsung.com.}
}


\maketitle
 
\begin{abstract}
In this paper, various Miller-compensated amplifiers are analyzed by using the two-port feedback analysis together with the root-locus diagram. The proposed analysis solves problems of Miller theorem/approximation that fail to predict a pole-splitting and that require an impractical assumption that an initial lower frequency pole before connecting a Miller capacitor in a two-stage amplifier should be associated with the input of the amplifier. Since the proposed analysis sheds light on how the closed-loop poles originate from the open-loop poles in the $s$-plane, it allows the association of the closed-loop poles with the circuit components and thus provides a design insight for frequency compensation. The circuits analyzed are two-stage Miller-compensated amplifiers with and without a current buffer and a three-stage nested Miller-compensated amplifier. 
\end{abstract}
\section{Introduction}
\begin{figure}[t]
\centering
\includegraphics[width=3.4in]{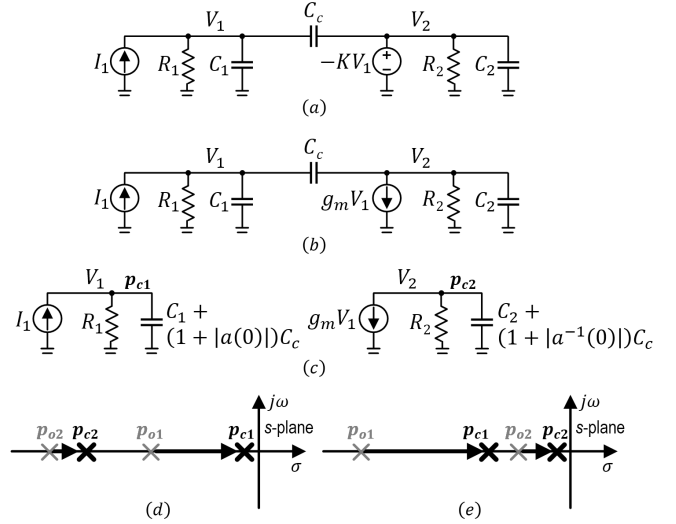}
\caption{Small-signal model of two-stage Miller-compensated amplifier using (a) controlled voltage source or (b) controlled current source. (c) Circuit obtained from applying Miller approximation to circuit in Fig. 1(b). Wrong estimation of the poles of circuit in Fig. 1(c) when the initial lower frequency pole before connecting $C_c$ is at (d) $V_1$ node or (e) $V_2$ node.}
\label{Fig1}
\end{figure}

\IEEEPARstart{M}{iller} theorem converts a feedback circuit into an open-loop circuit, allowing the association of the poles with the circuit nodes, which thus provides a design insight for frequency compensation. However, Miller theorem requires an impractical condition that the voltage-gain transfer function $a(s) = V_2/V_1$ in a given two-port network to be independent of $Y(s)$ which is an admittance connected across the ports~\cite{WHKi}. For example, consider a two-stage amplifier in Fig.~\ref{Fig1}(a) where $R_1$ and $C_1$ ($R_2$ and $C_2$) model the impedance at the input (output) and $C_c$ is a compensation capacitor. Since $Y(s) = sC_c$ and $a(s) = -K < 0 $ which is independent of $Y(s)$, the Miller theorem can be applied to this circuit. However, because a transistor is used to achieve the voltage-gain from the input to output in practice, the circuit should be modeled by using a controlled current source with a transconductance $g_m$ as shown in Fig.~\ref{Fig1}(b). Then, $a(s) = -g_mR_2(1-sC_c/g_m)/[1+s(C_2+C_c)R_2]$ is a function of $Y(s)$; thus, Miller theorem cannot be applied to this circuit. 

Alternatively, Miller approximation using a dc value of $a(s)$, $\abs{a(0)} = g_mR_2$, is often used to achieve the open-loop circuit as shown in Fig.~\ref{Fig1}(c). Assuming an initial lower frequency pole before connecting $C_c$ is associated with the input of the amplifier (i.e., $\abs{p_{o1}} = 1/R_1C_1 < \abs{p_{o2}} = 1/R_2C_2$), the Miller approximation predicts that the input pole moves to a lower frequency and is located such that $p_{c1} \simeq -1/g_mR_1R_2C_c$ which is true. However, it also predicts that a pole associated with the output becomes $p_{c2}$ with a lower frequency as well, which is in variance with a pole-splitting [See Fig.~\ref{Fig1}(d)]. Moreover, if the initial low-frequency pole is at the output (i.e., $\abs{p_{o1}} = 1/R_1C_1 > \abs{p_{o2}} = 1/R_2C_2$), which is a more general case because amplifiers are often required to drive a large load-capacitance, the Miller approximation predicts that the input pole moves to a lower frequency, which is completely false [See Fig.~\ref{Fig1}(e)]; the pole-splitting still occurs, implying the input pole moves to a higher frequency~\cite{WHKi}, which is very counter-intuitive.    

\begin{figure*}[t]
\centering
\includegraphics[width=6.8in]{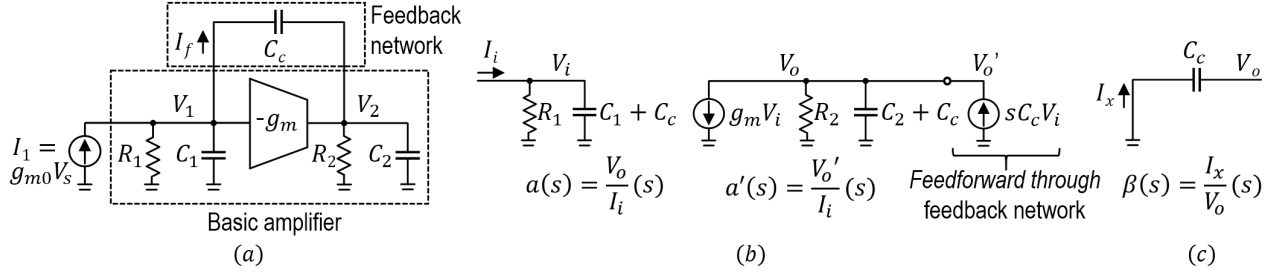}
\caption{(a) Two-stage Miller-compensated amplifier analyzed using two-port feedback analysis. (b) $a$-circuit. (c) $\beta$-circuit.}
\label{Fig2}
\end{figure*}

On the other hand, the direct analysis (i.e., solving node equations to achieve the desired transfer function) provides accurate the pole/zero locations~\cite{Leung_Analysis, Grasso2}. However, it is so complex that the design insight for frequency compensation cannot be readily obtained. 

In this paper, we use a two-port feedback analysis~\cite{Sedra, Gray} together with the root-locus diagram\cite{Franklin} to analyze two- and three-stage Miller-compensated amplifiers. This method solves the problems of Miller theorem/approximation mentioned above. Also, because the proposed analysis sheds light on how the closed-loop poles of the amplifier originated from the open-loop poles which can be readily found by inspection in the $s$-plane, it provides the design insight for frequency compensation.  

This paper is organized as follows. Section II analyzes a simple two-stage Miller-compensated amplifier using the proposed method. Specifically, we will see that a non-dominant pole location in a classical textbook~\cite{Sedra} should be modified to have a smaller magnitude when the right-half-plane (RHP) zero is neglected in the amplifier transfer function. Section III analyzes a two-stage Miller-compensated amplifier using a current buffer and shows that the stability can be improved due to the additional third-pole from the current buffer compared to the case without the current buffer. Section IV analyzes a three-stage nested Miller-compensated (NMC) amplifier~\cite{NMC}. Section~V concludes the paper. Finally, appendices are included to present a pole-splitting theorem and perform the direct analysis to investigate the exact locations of the non-dominant poles in the NMC amplifier.

\begin{figure}[t]
\centering
\includegraphics[width=3.4in]{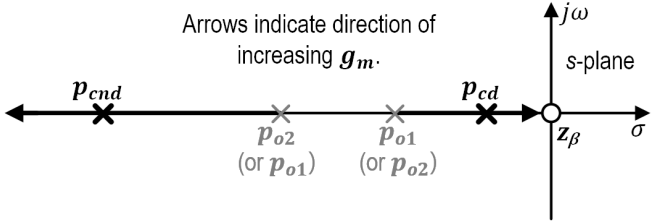}
\caption{Positive root-locus diagram for $a(s)\beta(s)$ given by~(\ref{LT1}). $p_{o1}$, $p_{o2}$ are the poles of $a(s)\beta(s)$ or equivalently the poles of $A(s)$ when the feedback loop is \textit{opened}. $p_{cd}$, $p_{cnd}$ are respectively the possible dominant and non-dominant poles of $A(s)$ when the feedback loop is \textit{closed}.}
\label{Fig3}
\end{figure}

\section{Two-Stage Miller-Compensated Amplifier} 
A two-stage Miller-compensated amplifier is shown in Fig.~\ref{Fig2}(a). This circuit is a feedback transimpedance amplifier that has a shunt-shunt feedback topology where the basic amplifier and the feedback network are shown in the dashed boxes. Thus, it can be decomposed into $a$- and $\beta$-circuit as shown Fig.~\ref{Fig2}(b) and (c), respectively. Note that the $a$-circuit includes the loading effect of the feedback network. Also, note that there are two open-loop transimpedances with [$a'(s)$] or without [$a(s)$] a non-inverting current source $sC_cV_i$ at the output that models the feedforward current through $C_c$.

Let us first investigate $a(s)$. By inspection of Fig.~\ref{Fig2}(b),
\begin{align}
a(s) & =  \frac{V_o}{I_{i}}(s) \nonumber \\ 
& =  -\left(R_1\parallel \frac{1}{s(C_1+C_c)}\right)g_m\left(R_2\parallel \frac{1}{s(C_2+C_c)}\right) \nonumber \\ 
& = -\frac{g_mR_1R_2}{[1+sR_1(C_1+C_c)][1+sR_2(C_2+C_c)]} \label{eq3} 
\end{align}
where $X\parallel Y = XY/(X+Y)$. Thus, $a(s)$ has two left-half-plane (LHP) real poles given by
\begin{align} 
p_{o1} & = -\frac{1}{R_1(C_1+C_c)}  \label{po1}  \\
p_{o2} &  = -\frac{1}{R_2(C_2+C_c)}    \label{po2}
\end{align}
where $p_{o1}$ and $p_{o2}$ are the open-loop poles associated with the input and output of the amplifier, respectively.

Next, $\beta(s)$ from Fig.~\ref{Fig2}(c) is given by
\begin{align}
\beta(s) = \frac{I_x}{V_o}(s) = -sC_c. \label{betafunc} 
\end{align}
Thus, $\beta(s)$ has a zero $z_{\beta}$ at the origin (i.e., $z_{\beta} = 0$).

\newcounter{mytempeqncnt}
\begin{figure*}[b]
\hrulefill
\vspace*{1pt}
\setcounter{mytempeqncnt}{\value{equation}} 
\setcounter{equation}{9}  
\begin{subequations} \label{ClTF3TF4}
\begin{align} 
A_{exact}(s)& =  \frac{-(g_m-sC_c)R_1R_2}{1 + s[R_1C_1 + R_2C_2 + (g_mR_1R_2 + R_1 + R_2)C_c] + s^2R_1R_2[C_1C_2 + C_c(C_1+C_2)]} \label{ClosedTF3}  \\ 
 & = \frac{-\left(R_1\parallel \frac{1}{s(C_1+C_c)}\right)(g_m-sC_c)\left(R_2\parallel \frac{1}{s(C_2+C_c)}\right)}{1+ \left(R_1\parallel \frac{1}{s(C_1+C_c)}\right)(g_m-sC_c)\left(R_2\parallel \frac{1}{s(C_2+C_c)}\right)sC_c} = \frac{a'(s)}{1+a'(s)\beta(s)}  \,\,\,\,\,\,\,\,\,\,\,\,\,\,\,\,\,\,\,\,\,\,\,\,\,\,\,\,\,\,\,\,\,\,\,\,\,\,\,\,\,\,\,  \label{ClosedTF4}
\end{align}
\end{subequations}
\begin{subequations} \label{ClTF1TF2}
\begin{align} 
 A(s) & =  \frac{-\left(R_1\parallel \frac{1}{s(C_1+C_c)}\right)g_m\left(R_2\parallel \frac{1}{s(C_2+C_c)}\right)}{1+ \left(R_1\parallel \frac{1}{s(C_1+C_c)}\right)g_m\left(R_2\parallel \frac{1}{s(C_2+C_c)}\right)sC_c} =  \frac{a(s)}{1+a(s)\beta(s)} \label{ClosedTF1}  \\ 
 & = \frac{-g_mR_1R_2}{1 + s[R_1C_1 + R_2C_2 + (g_mR_1R_2 + R_1 + R_2)C_c] + s^2R_1R_2[C_1C_2 + C_c(C_1+C_2) + \boldsymbol{C_c^2}]} \label{ClosedTF2} 
\end{align}
\end{subequations}
\setcounter{equation}{\value{mytempeqncnt}}
\end{figure*}

Combining (\ref{eq3}) and (\ref{betafunc}), the $a(s)\beta(s)$ is given by
\begin{align}
 a(s)\beta(s) = \frac{g_mR_1R_2C_cs}{[1+sR_1(C_1+C_c)][1+sR_2(C_2+C_c)]}. \label{LT1} 
\end{align}
Using~(\ref{LT1}), we can draw the positive root-locus diagram for increasing $g_m$ as shown in Fig.~\ref{Fig3} to find the poles of the closed-loop transimpedance $A(s) =V_2/I_1$ in Fig.~\ref{Fig2}(a). Note that $p_{o1}$, $p_{o2}$ are the poles of $a(s)\beta(s)$ or equivalently the poles of $A(s)$ when the feedback loop via $C_c$ is \textit{opened}, and $p_{cd}$, $p_{cnd}$ are the possible dominant and non-dominant poles of $A(s)$ when the feedback loop via $C_c$ is \textit{closed}. And $z_{\beta}$ is not a zero of $A(s)$ because it is from $\beta(s)$. Also note that the root-locus diagram includes the two cases where the initial lower frequency pole is either $p_{o1}$ or $p_{o2}$. As can be readily seen, the root-locus diagram reveals the fact that \textit{$p_{o1}$ and $p_{o2}$ always split apart for increasing $g_{m}$ irrespective of the condition about the initial low-frequency pole locations, which the Miller approximation fails to predict.}

Let us now find the exact locations of the closed-loop pole $p_{cd}$, $p_{cnd}$. Since $a(s)\beta(s)$ has two LHP real and distinct poles, one zero at the origin and a large midband value\footnote{The closed-loop poles are found from the midband value of $\abs{a(j\omega)\beta(j\omega)}$ if $a(s)\beta(s)$ has zero(s) at the origin~\cite{Roberge}.} $a_0\beta_0$ assuming $g_mR_1$, $g_mR_2$ and $C_c$ are large, we can apply the pole-splitting theorem presented in Appendix~A. Depending on whether the initial lower frequency pole of $a(s)\beta(s)$ is $p_{o1}$ or $p_{o2}$, $a_0\beta_0$ is expressed differently as
\begin{subnumcases}{a_0\beta_0 =}
   \frac{g_mR_2C_c}{C_1+C_c} & if $\abs{p_{o1}} < \abs{p_{o2}}$   \label{a0b0_1}
   \\
   \frac{g_mR_1C_c}{C_2+C_c} & if $\abs{p_{o1}} > \abs{p_{o2}}$. \label{a0b0_2}
\end{subnumcases}
Applying the pole-splitting relation~(\ref{PoleSplit}) yields
\begin{align}
p_{cd} & \simeq -\frac{1}{g_mR_1R_2C_c}  \label{pcd1}  \\
p_{cnd} &  \simeq -\frac{g_mC_c}{(C_1+C_c)(C_2+C_c)}.    \label{pcnd}
\end{align}

\begin{figure}[t]
\centering
\includegraphics[width=3.4in]{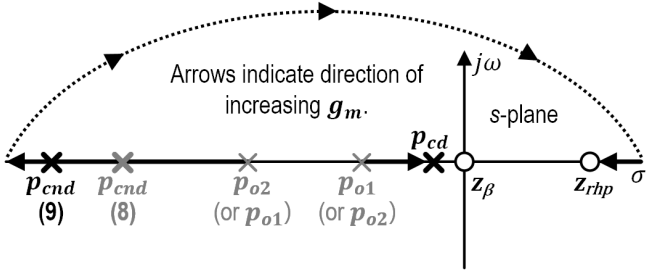}
\caption{Negative root-locus diagram for $a'(s)\beta(s)$.}
\label{Fig3-2}
\end{figure}

It should be mentioned that while the dominant pole $p_{cd}$ given by~(\ref{pcd1}) is the same as in~\cite{Sedra}, the non-dominant pole $p_{cnd}$ given by~(\ref{pcnd}) is in variance because 
\begin{equation}
p_{cnd} \,\, \textrm{in~\cite{Sedra}}  \simeq -\frac{g_mC_c}{(C_1+C_c)(C_2+C_c)-C_c^2},    \label{pcnd,sedra}
\end{equation}
implying (\ref{pcnd}) is located at a lower frequency than~(\ref{pcnd,sedra}). This variance is due to the assumption that the feedforward current through $C_c$ is neglected. This is verified as follow. The exact transimpedance $A_{exact}(s) = V_2/I_1$ of the two-stage Miller-compensated amplifier in Fig.~\ref{Fig2}(a) is given by~(\ref{ClosedTF3}). With some algebra, (\ref{ClosedTF3}) can be written in the form of the feedback equation as~(\ref{ClosedTF4}). Comparing~(\ref{ClosedTF4}) with (\ref{ClosedTF1}) obtained from the two-port feedback analysis shows that the only difference is that ($g_m - sC_c$) has been replaced by $g_m$. Thus, such a deviation vanishes if we include the effect of $sC_cV_i$ current source in Fig.~\ref{Fig2}(b), yielding
\addtocounter{equation}{2}
\begin{align}
& a'(s)  =  \frac{V_o'}{I_{i}}(s) \nonumber \\ 
& =  -\left(R_1\parallel \frac{1}{s(C_1+C_c)}\right)(g_m-sC_c)\left(R_2\parallel \frac{1}{s(C_2+C_c)}\right). \label{a'} 
\end{align}
Therefore, $A_{exact}(s) = a'(s)/[1+a'(s)\beta(s)]$ as shown in (\ref{ClosedTF4}), validating the two-port feedback analysis.

\begin{figure}[t]
\centering
\includegraphics[width=3.4in]{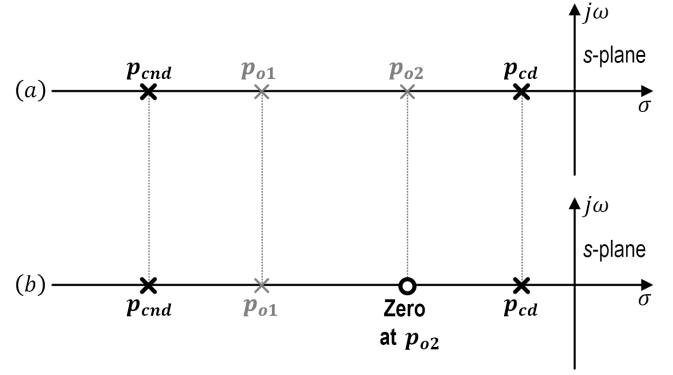}
\caption{Pole-zero diagram of (a) transimpedance $V_2/I_1$ and (b) input-impedance $V_1/I_1$ when $\abs{p_{o1}} > \abs{p_{o2}}$. }
\label{Fig4}
\end{figure}

\begin{figure*}[t]
\centering
\includegraphics[width=6.8in]{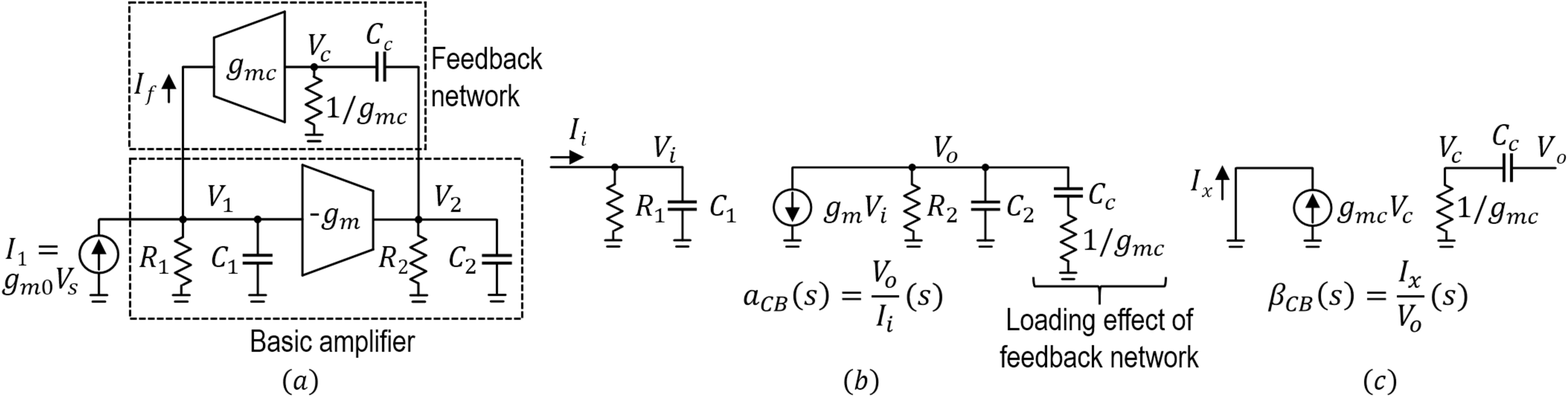}
\caption{(a) Two-stage Miller-compensated amplifier using current buffer analyzed using two-port feedback analysis. (b) $a$-circuit. (c) $\beta$-circuit.}
\label{Fig5}
\end{figure*}

\begin{figure}[t]
\centering
\includegraphics[width=3.4in]{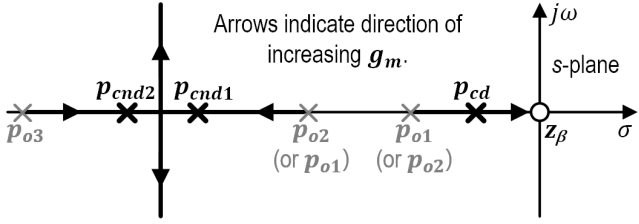}
\caption{Positive root-locus diagram for $a_{CB}(s)\beta_{CB}(s)$ given by~(\ref{LT1_CB}). $p_{o1}$, $p_{o2}$ and $p_{o3}$ are the poles of $a_{CB}(s)\beta_{CB}(s)$ or equivalently the poles of $A_{CB}(s)$ when the feedback loop is \textit{opened}. Also, $p_{cd}$, $p_{cnd1}$ and $p_{cnd2}$ are respectively the possible dominant, first and second non-dominant poles of $A_{CB}(s)$ when the feedback loop is \textit{closed}.}
\label{Fig6}
\end{figure}

Indeed, \textit{the non-dominant pole given by (\ref{pcnd}) is more accurate than (\ref{pcnd,sedra}) if the RHP zero ($z_{rhp} = g_m/C_c$) of the two-stage Miller-compensated amplifier is assumed to be at high frequency so as to be neglected.} $z_{rhp}$ can be ignored by assuming a large $g_m$ so that $(g_m - sC_c) \simeq g_m$ at the frequencies of interest. However, because the term $(g_m - sC_c)$ also exists in the denominator of (\ref{ClosedTF4}), such condition also alter the non-dominant pole location by modifying the coefficient of $s^2R_1R_2$ to include an additional term of $\boldsymbol{C_c^2}$ as shown in the denominator of~(\ref{ClosedTF2}), which results in~(\ref{pcnd}) to have a smaller magnitude than~(\ref{pcnd,sedra}).

Such increase in the magnitude of the non-dominant pole can also be explained using the root-locus diagram for $a'(s)\beta(s)$     as shown in Fig.~\ref{Fig3-2}. Note that the negative locus rule~\cite{Franklin} is applied here because of the low-frequency sign reversal associated with the RHP zero~\cite{Roberge}. Similar to Fig.~\ref{Fig3}, $p_{cnd}$ moves toward high frequency in the LHP as $g_m$ increases. However, it eventually moves into the RHP, manifesting itself at $s = z_{rhp}^+$ if $g_m \rightarrow \infty$; thus, for a given finite $g_m$, $p_{cnd}$ is located at nearer to $s= -\infty$ when the RHP zero is included as illustrated in Fig.~\ref{Fig3-2}.

So far, we have investigated the pole/zero locations of the transimpedance $V_2/I_1$ of the circuit in Fig.~\ref{Fig2}(a). It is worth mentioning of the pole/zero locations of the input-impedance $V_1/I_1$. Since the natural structure of the circuit is unchanged, $p_{cd}$, $p_{cnd}$ are also unchanged. However, the voltage-amplifier $V_o/V_i$ or $V'_o/V_i$ in Fig.~\ref{Fig2}(b) that has a pole $p_{o2}$ is now included \textit{not} in the basic amplifier \textit{but} in the feedback network, and thus $p_{o2}$ becomes a zero in the input-impedance. For example, the pole-zero diagram of the transimpedance $V_2/I_1$, input-impedance $V_1/I_1$ when $\abs{p_{o1}} > \abs{p_{o2}}$ are shown in Fig.~\ref{Fig4}(a), (b), respectively. The pole/zero pattern of the input-impedance can be employed as a lag compensating network to control a damping factor of the complex poles in a three-stage amplifier in \cite{DFCFC}.

\section{Two-Stage Miller-Compensated Amplifier Using Current Buffer} 
The RHP zero in the Miller-compensated two-stage amplifier in the previous section was ignored by assuming a large $g_m$, which requires a large power consumption in practice. Instead, the RHP zero can be removed by employing a unilateral current buffer~\cite{Ahuja, Hurst} in the feedback network, allowing only a current through $C_c$ to flow backward from the output to the input of the transconductor ($g_m$) as shown in Fig.~\ref{Fig5}(a). Note that the feedback network now has a current buffer with an input impedance of $1/g_{mc}$ and the transconductance $g_{mc}$. Since this circuit also has a shunt-shunt feedback topology, it can be decomposed into $a$- and $\beta$-circuit as shown Fig.~\ref{Fig5}(b) and (c), respectively. Note that unlike the amplifier without the current buffer, the feedback loading only occurs at the output of the amplifier in the $a$-circuit. 

Assuming $g_{mc}R_2 \gg 1$, the open-loop transimpedance $a_{CB}(s)$ is 
\begin{align}
& a_{CB}(s)  =  \frac{V_o}{I_{i}}(s) \nonumber \\ 
& = -\frac{g_mR_1R_2(1+sC_c/g_{mc})}{[1+sR_1C_1][1+sR_2(C_2+C_c)]\left[1+s\frac{C_2\parallel C_c}{g_{mc}}\right]}. \label{eq3_CB} 
\end{align}
Thus, $a_{CB}(s)$ has three LHP real poles as
\begin{align} 
p_{o1} & = -\frac{1}{R_1C_1}  \label{po1_CB}  \\
p_{o2} &  = -\frac{1}{R_2(C_2+C_c)}    \label{po2_CB} \\
p_{o3} &  = -\frac{g_{mc}}{C_2 \parallel C_c}    \label{po3_CB}
\end{align}
and one LHP zero as 
\begin{align} 
z_{a} & = -\frac{g_{mc}}{C_c}  \label{za_CB} 
\end{align}

From Fig.~\ref{Fig5}(c), $\beta_{CB}(s)$ is 
\begin{align}
\beta_{CB}(s) = \frac{I_x}{V_o}(s) = -\frac{sC_c}{1+sC_c/g_{mc}}. \label{betafunc_CB} 
\end{align}
Thus, $\beta_{CB}(s)$ has one LHP real pole as
\begin{align}
p_{\beta} = -\frac{g_{mc}}{C_c} \label{pbeta_CB} 
\end{align}
and a zero $z_{\beta} = 0$. 

\begin{figure*}[b]
\hrulefill
\vspace*{1pt}
\setcounter{mytempeqncnt}{\value{equation}} 
\setcounter{equation}{23}  
\begin{align} 
A_{CB}(s)    = \frac{a_{CB}(s)}{1+a_{CB}(s)\beta_{CB}(s)}   \simeq \frac{-g_mR_1R_2 \left(1+s\dfrac{C_c}{g_{mc}}\right)}{(1+sg_mR_1R_2C_c)\left(1+s\dfrac{C_cC_2 + g_{mc}R_1C_1(C_c+C_2)}{g_mR_1g_{mc}C_c}+s^2\dfrac{C_1C_2}{g_mg_{mc}} \right)} \label{ACB}
\end{align}
\setcounter{equation}{\value{mytempeqncnt}}
\end{figure*}

\begin{figure}[t]
\centering
\includegraphics[width=3.4in]{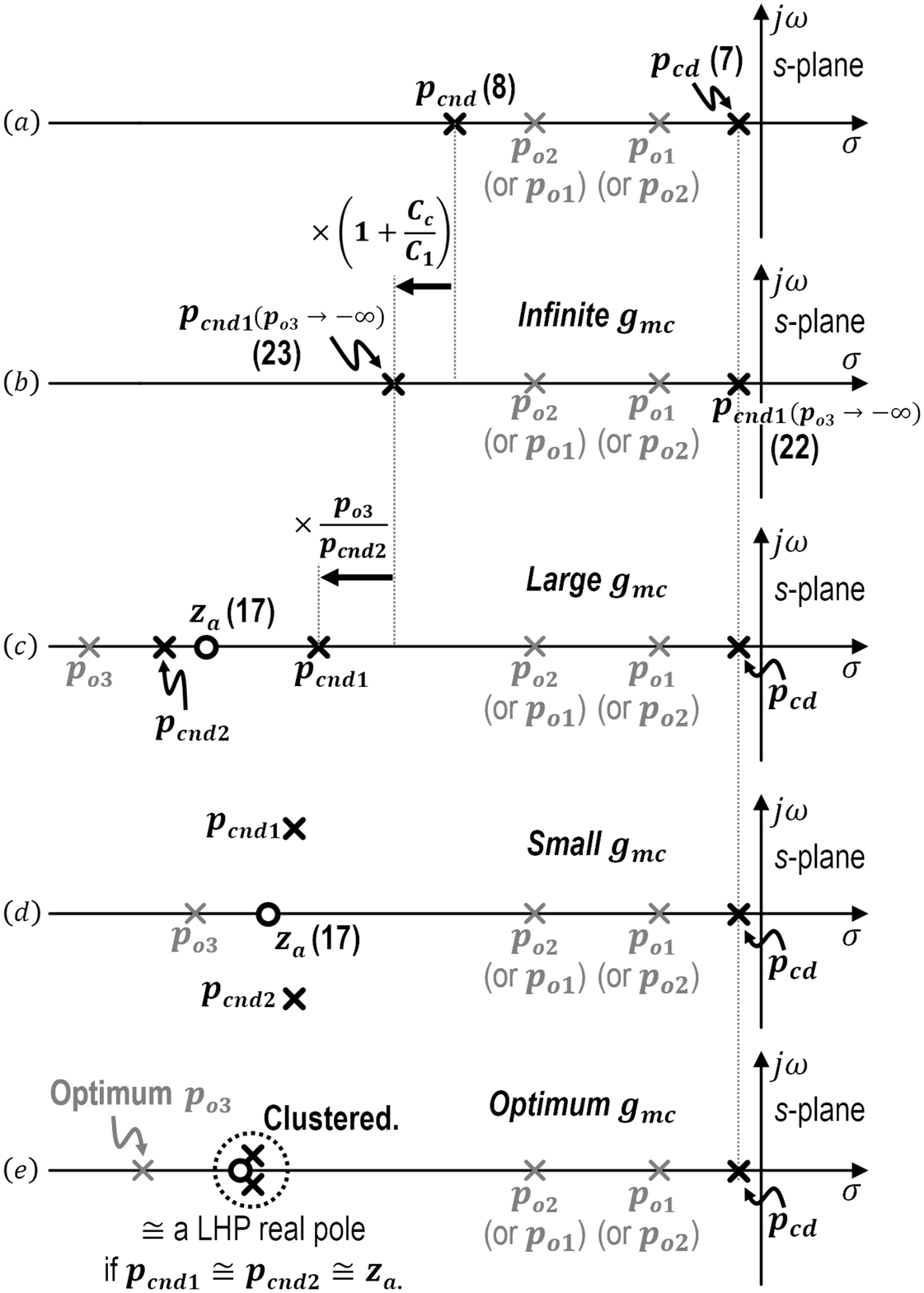}
\caption{Pole-zero diagram of the closed-loop transimpedance $V_2/I_1$ of the two-stage amplifier (Not to scale): (a) without current buffer, (b) with ideal current buffer that has infinite $g_{mc}$, (c) with non-ideal current buffer that has large $g_{mc}$, (d) with non-ideal current buffer that has small $g_{mc}$, (e) with non-ideal current buffer that has optimum $g_{mc}$.}
\label{Fig7}
\end{figure}

Since (\ref{pbeta_CB}) coincides with (\ref{za_CB}), the $a_{CB}(s)\beta_{CB}(s)$ has the three LHP poles, a zero $z_{\beta}$ at the origin, and is expressed as
\begin{align}
&  a_{CB}(s)\beta_{CB}(s) \nonumber \\ 
& = \frac{g_mR_1R_2C_cs}{[1+sR_1C_1][1+sR_2(C_2+C_c)]\left[1+s\frac{C_2\parallel C_c}{g_{mc}}\right]} . \label{LT1_CB} 
\end{align}
The positive root-locus diagram of $a_{CB}(s)\beta_{CB}(s)$ for increasing $g_m$ is shown in Fig.~\ref{Fig6}. The resulting loci indicate the locations of the poles of the closed-loop transimpedance $A_{CB}(s) = V_2/I_1$ in Fig.~\ref{Fig5}(a). As $g_{m}$ increases, $p_{o1}$ and $p_{o2}$ split apart and become the dominant pole $p_{cd}$ and the first non-dominant pole $p_{cnd1}$, respectively. Also, $p_{o3}$ moves toward lower frequency and becomes the second non-dominant pole $p_{cnd2}$ for increasing $g_{m}$, implying that $p_{cnd1}$ and $p_{cnd2}$ can form a complex pole pair beyond some value of $g_{m}$. Thus, we should analyze how effective this amplifier is compared to the previous amplifier in Section II and also investigate the effect of the third pole $p_{o3}$ on the stability of the amplifier.

\begin{figure*}[t]
\centering
\includegraphics[width=7.2in]{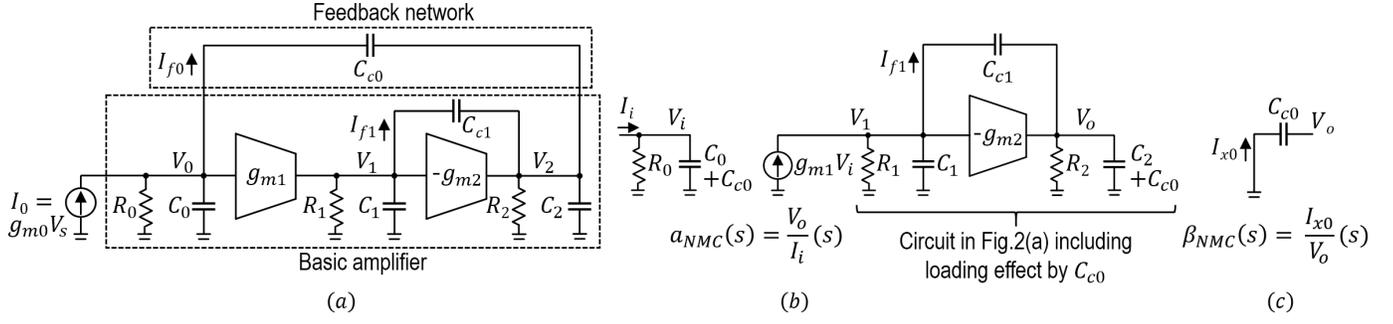}
\caption{(a) Three-stage NMC amplifier analyzed using two-port feedback analysis. (b) $a$-circuit. (c) $\beta$-circuit.}
\label{Fig8}
\end{figure*}

First, let us compare the Miller compensated amplifier using the current buffer with the previous amplifier by assuming that $p_{o3} \rightarrow -\infty$ so as to be neglected. It is achieved by assuming the ideal current buffer with zero input impednace (i.e., $g_{mc} \rightarrow \infty$). Then, (\ref{LT1_CB}) reduces to have the two LHP poles, a zero at the origin and the midband value $a_{0}\beta_{0}|_{CB}$ as 
\begin{subnumcases}{a_0\beta_0|_{CB} =}
   \frac{g_mR_2C_c}{C_1} & if $\abs{p_{o1}} < \abs{p_{o2}}$ \label{a0b0_CB1}
   \\
   \frac{g_mR_1C_c}{C_2+C_c} & if $\abs{p_{o1}} > \abs{p_{o2}}$. \label{a0b0_CB2}
\end{subnumcases}
Since the midband value is typically very large, we apply the pole-splitting theorem in Appendix A to find the splitted closed-loop poles as
\begin{align}
p_{cd (p_{o3} \rightarrow-\infty)} & \simeq -\frac{1}{g_mR_1R_2C_c}  \label{pcd1_CB} \\
p_{cnd1 (p_{o3} \rightarrow-\infty)} &\simeq -\frac{g_m}{C_2+C_c} \frac{C_c}
{C_1}.     \label{pcnd1_CB}
\end{align}

While the dominant pole (\ref{pcd1_CB}) is the same with (\ref{pcd1}), the non-dominant pole given by (\ref{pcnd1_CB}) is located at higher frequency than (\ref{pcnd}) by a multiplication factor of $1+C_c/C_1$ which is typically large because $C_1$ is a small parasitic capacitance. Thus, the amplifier with the current buffer has the better relative stability than the amplifier without the current buffer if their gain-bandwidth products (GBWs) are the same.\footnote{GBWs of the voltage-gain $V_{2}/V_s$ of the amplifier in Fig.~\ref{Fig2}(a) and Fig.~\ref{Fig5}(a) are expressed as $g_{m0}/C_{c}$ by replacing the input current source $I_1$ with $g_{m0}V_s$ where $g_{m0}$ is an additional input transconductance applied to the input voltage $V_s$.} The pole-zero diagrams of the transimpedance $V_2/I_1$ of the two-stage amplifier without the current buffer and the two-stage amplifier with the ideal current buffer is shown in Fig.~\ref{Fig7}(a), (b), respectively. 

\addtocounter{equation}{1}
\begin{figure*}[b]
\hrulefill
\vspace*{1pt}
\setcounter{mytempeqncnt}{\value{equation}} 
\setcounter{equation}{26}   
\begin{align}  
a_{NMC}(s)\beta_{NMC}(s)  =  \frac{g_{m1}g_{m2}R_0R_1R_2C_{c0}s}{[1+sR_0(C_0+C_{c0})][1+sg_{m2}R_1R_2C_{c1}]\left[1+s\dfrac{(C_1+C_{c1})(C_2+C_{c0}+C_{c1})}{g_{m2}C_{c1}}\right]}  \label{abetaNMC} 
\end{align}
\setcounter{equation}{\value{mytempeqncnt}}
\end{figure*}

Such increase of magnitude of the non-dominant pole is explained as follows. When $\abs{p_{o1}} < \abs{p_{o2}}$, (\ref{a0b0_CB1}) is larger than (\ref{a0b0_1}) by the multiplication factor of $1+C_c/C_1$; thus, such increased midband value help move $p_{o2}$ given by~(\ref{po2_CB}) to become $p_{cnd1}$ that is located at higher frequency than (\ref{pcnd}). When $\abs{p_{o1}} > \abs{p_{o2}}$, though the midband value given by (\ref{a0b0_CB2}) is unchanged compared to (\ref{a0b0_2}), the initial higher frequency pole $p_{o1}$ given by (\ref{po1_CB}) is located at more higher frequency than (\ref{po1}) by the multiplication factor of $1+C_c/C_1$ because the loading effect does not occur at the input of the amplifier; thus, $p_{cnd1}$ is located at higher frequency than (\ref{pcnd}) as well.

Next, we consider a non-ideal current buffer that has a finite $g_{mc}$ so that $p_{o3}$ and thus $p_{cnd2}$ are not neglected.  Assuming $g_{m}, g_{mc} \gg 1/R_1, 1/R_2$; $C_c, C_2 \gg C_1$; $g_{m}R_1 \gg C_2 / C_c$, $A_{CB}(s)$ is given by~(\ref{ACB}) at the bottom of the page. Note that the dominant pole $p_{cd}$ is the same as~(\ref{pcd1_CB}); this is because $p_{o1}$ is typically located at much lower frequency than $p_{o3}$ and thus $\abs{a_{CB}(j\omega)\beta_{CB}(j\omega)}$ at $\omega = \abs{p_{o1}}$ is little affected by $p_{o3}$. Therefore, $p_{o1}$ moves toward the origin for the same amount and becomes $p_{cd}$ as before when feedback-loop is closed for a given $g_m$.

The other poles $p_{cnd1}$, $p_{cnd2}$ can be found by factoring the second-order polynomial in the denominator of $A_{CB}(s)$. Though simple expressions for $p_{cnd1}$, $p_{cnd2}$ cannot be readily generated, they have the following simple relation: 
\begin{align}
  p_{cnd1}p_{cnd2}  =  \frac{g_{m}g_{mc}}{C_1C_2}.  \label{pcnd_product}  
\end{align}
Since the right-hand side of (\ref{pcnd_product}) can be obtained by multiplying~(\ref{po3_CB}) with (\ref{pcnd1_CB}), we write
\begin{align}
p_{cnd1} = \frac{p_{o3}}{p_{cnd2}} p_{cnd1(p_{o3} \rightarrow-\infty)}.   \label{pcnd1}  
\end{align}
Thus, $p_{cnd1}$ is modified with a multiplication factor of $p_{o3}/p_{cnd2}$ compared to $p_{cnd1(p_{o3} \rightarrow-\infty)}$. Let us consider the following two cases where $g_{mc}$ and thus $p_{o3}$~(\ref{po3_CB}) are different but the midband values of $a_{CB}(s)\beta_{CB}(s)$ are preserved since (\ref{a0b0_CB1}) and (\ref{a0b0_CB2}) are not functions of $g_{mc}$: 1) $g_{mc}$ is large and thus $p_{o3}$ is located at high frequency so that $p_{cnd1}$ and $p_{cnd2}$ are real and 2) $g_{mc}$ is small and thus $p_{o3}$ is located at low frequency so that $p_{cnd1}$ and $p_{cnd2}$ are complex.

1) When $g_{mc}$ is large so that $p_{cnd1}$ and $p_{cnd2}$ are real, $p_{o3}/p_{cnd2}$ is obviously larger than "1" as shown in the root-locus diagram in Fig.~\ref{Fig6}; thus, $p_{cnd1}$ is located at higher frequency than $p_{cnd1(p_{o3} \rightarrow-\infty)}$ \textit{as if attracted by} $p_{cnd2}$ as shown in the pole-zero diagram of $V_2/I_1$ in Fig.~\ref{Fig7}(c). Note that because $z_a$~(\ref{za_CB}) is the zero of $a_{CB}(s)$ and $p_{\beta}$~(\ref{pbeta_CB}) is not the pole of $a_{CB}(s)\beta_{CB}(s)$, a zero $z_a$ exists in $A_{CB}(s)$. Though $p_{cnd2}$ is at the lower frequency than in the previous ideal current buffer case, the stability is not degraded if $z_a$ and $p_{cnd2}$ are closely spaced so as to be canceled as shown in Fig.~\ref{Fig7}(c). Thus, this case is better than the Miller compensation with the ideal current buffer since it achieves both the improved stability margin and the lower power consumption with the finite $g_{mc}$. 

2) When $g_{mc}$ is small so that $p_{cnd1}$ and $p_{cnd2}$ are complex, the magnitude of the complex poles is still larger than that of $p_{cnd1(p_{o3} \rightarrow-\infty)}$~\cite{Myungjun}. Thus, power consumption can be further reduced by decreasing $g_{mc}$ compared to the previous real non-dominant poles case without deteriorating the stability. However, since a real part of the complex pole-pair is fixed at $s \simeq p_{o3}/2$, too small $g_{mc}$ can result in a small damping ratio of the pole-pair, degrading stability margins shown in Fig.~\ref{Fig7}(d). Therefore, for a given $g_m$, an optimum design that achieves both good stability and low power consumption can be achieved by locating $p_{o3}$ with the optimum $g_{mc}$, allowing $p_{cnd1}$ and $p_{cnd2}$ to be located at  near $z_a$ so that they can be considered a LHP real pole as shown in Fig.~\ref{Fig7}(e) (i.e., $p_{c2} \simeq p_{cM} \simeq z_a$).

\begin{figure*}[b]
\hrulefill
\vspace*{1pt}
\setcounter{mytempeqncnt}{\value{equation}} 
\setcounter{equation}{34}  
\begin{equation} 
A_{NMC}(s) = \frac{V_2}{I_0}(s) = \frac{a_{NMC}(s)}{1+a_{NMC}(s)\beta_{NMC}(s)}  \simeq  \frac{-g_{m1}g_{m2}R_0R_1R_2}{(1+sg_{m1}g_{m2}R_0R_1R_2C_{c0})\left[1+s\dfrac{C_{c1}}{g_{m1}}+s^2\dfrac{C_{c1}C_2}{g_{m1}g_{m2}} \right]}  \label{ANMC}
\end{equation}
\setcounter{equation}{\value{mytempeqncnt}}
\end{figure*}

\section{Three-Stage NMC Amplifier} 
In this section, we analyze three-stage NMC amplifiers~\cite{NMC} by using two-port feedback analysis. 

A three-stage NMC amplifier is shown in Fig.~\ref{Fig8}(a). Note that two Miller capacitors $C_{c0}$, $C_{c1}$ are employed and the input current source $I_0$ is expressed as $g_{m0}V_s$ where $V_s$ is the input voltage of the NMC amplifier. Also, note that a two-stage Miller-compensated amplifier in Section II is included in this NMC amplifier. This circuit is also a feedback transimpedance amplifier that has a shunt-shunt feedback topology where the basic amplifier and the feedback network are shown in the dashed boxes in Fig.~\ref{Fig8}(a). Thus, it can be decomposed into $a$- and $\beta$-circuit as shown Fig.~\ref{Fig8}(b) and (c), respectively. It should be noted that the $a$-circuit is drawn by neglecting the signal feedforward through $C_{c0}$ for simplicity. Also, the $a$-circuit includes the two-stage amplifier in Fig.~\ref{Fig2}(a) where the output impedance includes the loading effect by $C_{c0}$. 

Using the same approches and the results in Section II, $a_{NMC}(s)\beta_{NMC}(s)$ which is given by~(\ref{abetaNMC}) shown at the bottom of the page has three LHP poles given by
\addtocounter{equation}{1}
\begin{align} 
p_{o0} & = -\frac{1}{R_0(C_0+C_{c0})} \simeq -\frac{1}{R_0C_{c0}}    \label{po0_nmc}  \\
p_{o1} &  = -\frac{1}{g_{m2}R_1R_2C_{c1}}    \label{po1_nmc} \\ 
p_{o2} &  = -\frac{g_{m2}C_{c1}}{(C_1+C_{c1})(C_2+C_{c0}+C_{c1})} \simeq  -\frac{g_{m2}}{C_2}    \label{po2_nmc}
\end{align}
where the approximations in (\ref{po0_nmc}) and (\ref{po2_nmc}) follow if $C_2 \gg  C_{c0}, C_{c1} \gg C_0, C_1$ since typically $C_2$ is a large load capacitance and $C_0, C_1$ are small parasitics. Also, $a_{NMC}(s)\beta_{NMC}(s)$ has a $z_{\beta}$ at the origin from the $\beta_{NMC}(s) = I_{f0}/V_o$ in Fig.~\ref{Fig8}(c). Note that $p_{o1}$ and $p_{o2}$ are expressed by neglecting the signal feedforward through $C_{c1}$.

\begin{figure}[t]
\centering
\includegraphics[width=3.4in]{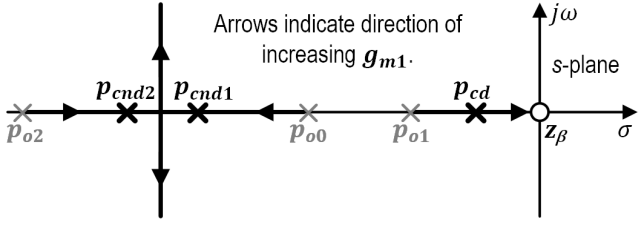}
\caption{Positive root-locus diagram for $a_{NMC}(s)\beta_{NMC}(s)$ given by~(\ref{abetaNMC}). $p_{o0}$, $p_{o1}$ and $p_{o2}$ are the poles of $a_{NMC}(s)\beta_{NMC}(s)$ or equivalently the poles of $A_{NMC}(s) = V_2/I_0$ when the feedback loop via $C_{c0}$ is \textit{opened}. Also, $p_{cd}$, $p_{cnd1}$ and $p_{cnd2}$ are respectively the possible dominant, first and second non-dominant poles of $A_{NMC}(s)$ when the feedback loop via $C_{c0}$ is \textit{closed}.}
\label{Fig10}
\end{figure}

Typically, $\abs{p_{o1}} < \abs{p_{o0}} < \abs{p_{o2}}$ because of a large $g_{m2}$ in the design of the NMC amplifier~\cite{Leung_Analysis}. Thus, the positive root-locus diagram for increasing $g_{m1}$ can be drawn as shown in Fig.~\ref{Fig10}. As can be seen, the two low frequency poles $p_{o1}$ and $p_{o0}$ split apart as $g_{m1}$ increases and become a dominant pole $p_{cd}$ and a first non-dominant pole $p_{cnd1}$, respectively. It should also be noted that $p_{o2}$ moves toward low frequency and becomes a second non-dominant pole $p_{cnd2}$ as $g_{m1}$ increases. Also, note that $p_{cnd1}$ and $p_{cnd2}$ can form a complex-conjugate pole-pair beyond some value of $g_{m1}$. 

\begin{figure}[t]
\centering
\includegraphics[width=3.4in]{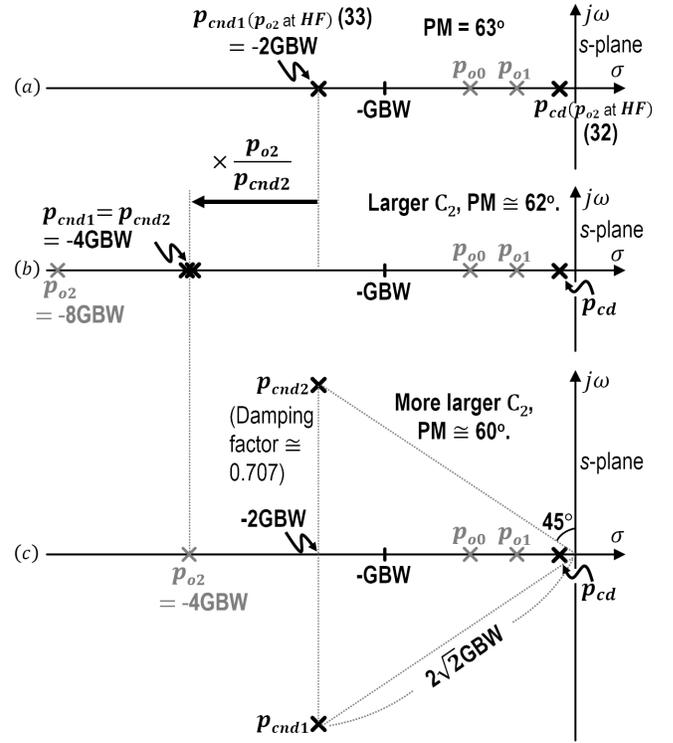}
\caption{Pole-zero diagram of the voltage-gain $V_2/V_s$ of the three-stage NMC amplifier (Not to scale): (a) when $p_{o2}$ is located at a very high frequency so as to be neglected, (b) when $p_{o2}$ is located at $s=-8$GBW, and (c) when $p_{o2}$ is located at $s=-4$GBW.}
\label{Fig11}
\end{figure}

Using a similar approach in the Section III, let us first assume $p_{o2}$ is located at very high frequency so as to be neglected and investigate a dominant ($p_{cd}$) and a first non-dominant ($p_{cnd1}$) pole locations; in this case, (\ref{abetaNMC}) can be approximated to have the two LHP poles, a zero at the origin, and a midband value $a_{0}\beta_{0}|_{NMC}$ as given by  
\begin{equation}
a_0\beta_0|_{NMC} = \frac{g_{m1}R_0C_{c0}}{C_{c1}}  \label{a0b0_NMC}
\end{equation}
which is typically large. Thus, the splitted closed-loop poles can be found by applying the pole-splitting relation~(\ref{PoleSplit}), which results in
\begin{align}
p_{cd (p_{o2} \, \textrm{at} \, HF)} & \simeq -\frac{1}{g_{m1}g_{m2}R_0R_1R_2C_{c0}}  \label{pcd_NMC} \\
p_{cnd1 (p_{o2} \, \textrm{at} \, HF)} & \simeq  -\frac{g_{m1}}{C_{c1}} \label{pcnd1_NMC} 
\end{align}
Thus, GBW of this NMC amplifier is $g_{m0}/C_{c0}$. Typically, a magnitude of the first non-dominant pole given by~(\ref{pcnd1_NMC}) is set to be
\begin{align}
\frac{g_{m1}}{C_{c1}} = 2\textrm{GBW} = 2\frac{g_{m0}}{C_{c0}}  \label{GBW1}
\end{align}
so that the phase margin (PM) $\simeq$ 90\textdegree{}$-tan^{-1}(1/2) \simeq 63$\textdegree{}~\cite{Leung_Analysis}; the pole-zero diagram of the voltage-gain of this NMC amplifier is illustrated in~Fig.~\ref{Fig11}(a). It is worth mentioning that the proposed analysis shows that $p_{o2}$ should be located at a high frequency such that $\abs{p_{o2}} = \abs{p_{cnd2}} \gg $ GBW to satisfy (\ref{GBW1}), implying a seperate pole approach in \cite{Leung_Analysis} that satisfies not only~(\ref{GBW1}) but $\abs{p_{cnd2}}$ = 2 GBW cannot be achieved.

Next, we consider the case when $p_{o2}$ is located at a lower frequency and investigate its effect on the stability. Though this can be achieved by decreasing $g_{m2}$, it will also affect $p_{o1}$~given by~(\ref{po1_nmc}). Instead, increasing $C_2$ only allows $p_{o2}$ to be located at a lower frequency. Then, as mentioned eariler, $p_{o2}$ moves toward low frequency and becomes the second-nondominant pole $p_{cnd2}$. Similar to the case in the Miller compensated two-stage using the non-ideal current buffer in Section III, the first non-dominant pole $p_{cnd1}$ here is at a higher frequency than (\ref{pcnd1_NMC}) as if attracted by $p_{cnd2}$. This is verified as follows.

Assuming $g_{m1}, g_{m2} \gg 1/R_0, 1/R_1, 1/R_2$; $C_2 \gg C_{c0}, C_{c1} \gg C_0, C_1$; $A_{NMC}(s) =  V_2 / I_0$ in Fig.~\ref{Fig8}(a) is approximately expressed as~(\ref{ANMC}) shown at the bottom of the page. From a quadratic function in the denominator of~(\ref{ANMC}), $p_{cnd1}$ and $p_{cnd2}$ have the following relation:  
\addtocounter{equation}{1}
\begin{align}
p_{cnd1}p_{cnd2}  \simeq  \frac{g_{m1}g_{m2}}{C_{c1}C_2}.  \label{pcnd_product_NMC}  
\end{align}
Because the right-hand side of~(\ref{pcnd_product_NMC}) can also be obtained by multiplying~(\ref{po2_nmc}) with (\ref{pcnd1_NMC}), we can write
\begin{align}
p_{cnd1} \simeq \frac{p_{o2}}{p_{cnd2}} p_{cnd1 (p_{o2} \, \textrm{at} \, HF)}.   \label{pcnd1_NMC2}  
\end{align}
Thus, $p_{cnd1}$ is located at higher frequency than $p_{cnd1 (p_{o2} \, \textrm{at} \, HF)}$ by a multiplication factor of $p_{o2}/p_{cnd2}$ that is obviously larger than unity as shown in the root-locus diagram in Fig.~\ref{Fig10}. This implies that \textit{even if $p_{o2}$ is not located at very high frequency, the stability margin may be little affected by $p_{cnd2}$ because the magnitude of $p_{cnd1}$ is also increased.}

For example, we consider two cases that have different values of $C_2$ so that 1) $p_{o2} = -8$GBW, and 2) $p_{o2} = -4$GBW and the condition~(\ref{GBW1}) holds in the both cases. The exact analysis is carried out for the both cases in Appendix~B. 

1) When $p_{o2} = -8$GBW, the NMC amplifier has the pole-zero diagram as illustrated in~Fig.~\ref{Fig11}(b). Note that the dominant pole $p_{cd}$ is at the same location as in the previous case because $p_{o1}$ is typically located at much lower frequency than $p_{o2}$ and thus $\abs{a_{NMC}(j\omega)\beta_{NMC}(j\omega)}$ at $\omega = \abs{p_{o2}}$ is little affected by $p_{o2}$. However, $p_{o0}$ and $p_{o2}$ attract each other, becoming the first~($p_{cnd1}$) and second non-dominant~($p_{cnd2}$) pole, and form a double pole-pair at $s = -4$GBW, which results in PM $\simeq$ 90\textdegree{}$-2tan^{-1}(1/4) \simeq 62$\textdegree{}. Thus, the stability of the NMC amplifier is little affected compared to the previous case.

2) When $p_{o2} = -4$GBW, the NMC amplifier has the pole-zero diagram as illustrated in~Fig.~\ref{Fig11}(c). The location of $p_{cd}$ is still the same as the previous cases. However, $p_{cnd1}$ and $p_{cnd2}$ form a complex-conjugate pole-pair that has the same magnitude of the real- and imaginary-part as 2GBW (i.e.,  $\abs{p_{cnd1}}=\abs{p_{cnd2}}=2\sqrt{2}$GBW) with a damping ratio ($\xi$) of $1/\sqrt{2} \simeq 0.707$, which results in 
\begin{align}
PM \simeq 90\textrm{\textdegree{}} - tan^{-1}\left[\frac{2\xi \frac{GBW}{2\sqrt{2}GBW}}{1-(\frac{GBW}{2\sqrt{2}GBW})^2}\right] \simeq 60\textrm{\textdegree{}}.    \label{PM1}  
\end{align}
Thus, the stability of the NMC amplifier is also little affected compared to the previous cases.
\section{Conclusion} 
In this paper, various Miller-compensated amplifiers are analyzed by using the two-port feedback analysis together with the root-locus diagram. This method solves the problems of Miller theorem/approximation that fail to predict the pole-splitting and that requires an  an impractical assumption where an initial low-frequency pole before connecting a Miller capacitor in a multi-stage amplifier should be associated with the input of the amplifier. Also, because the proposed analysis sheds light on how the closed-loop poles of the amplifier originated from the open-loop poles which can be readily found by inspection in the $s$-plane, it provides the design insight for frequency compensation, helping allow the association of the closed-loop poles with the circuit parameters. 

Specifically, in Section II, the proposed analysis shows that when the RHP zero is neglected the nondominant pole location of the two-stage Miller-compensated amplifier should be modified to have a smaller magnitude than in a classical textbook\cite{Sedra}. Also, in Section III, when the current buffer is used in a Miller compensated two-stage amplifier to block the feedforward path through the compensation capacitor, the stability can be improved because of the absence of the RHP zero and the loading effect of compensation capacitor at the input of the amplifier. Moreover, the additional third pole from the current buffer and the zero of the $a$-circuit can be optimally located to improve the stability and to lower the power consumption. Finally, in Section IV, the stabiltiy of the three-stage NMC amplifier can be little affected even if the second non-dominant pole is considered because the first non-dominant pole is located at a higher frequency as if attracted by the second non-dominant pole.
\appendices
\section{Pole-splitting Theorem}
\textit{Let $a(s)\beta(s)$ of a feedback circuit has two distinct real poles, $p_{od}$ and $p_{ond}$, in the left-half of the $s$-plane such that $\abs{p_{od}} < \abs{p_{ond}}$, a zero at the origin, and a midband value $a_0\beta_0 \gg 2$. Then, the following Pole-splitting relation holds:
\begin{equation}
 \boxed{  a_0\beta_0  \simeq \frac{p_{od}}{p_{cd}}  \simeq \frac{p_{cnd}}{p_{ond}}} \label{PoleSplit}
\end{equation}
where $p_{cd}$ and $p_{cnd}$ are the closed-loop poles of the circuit.}  
\begin{flushright}
$\blacksquare$
\end{flushright}

\begin{figure}[t]
\centering
\includegraphics[width=3.4in]{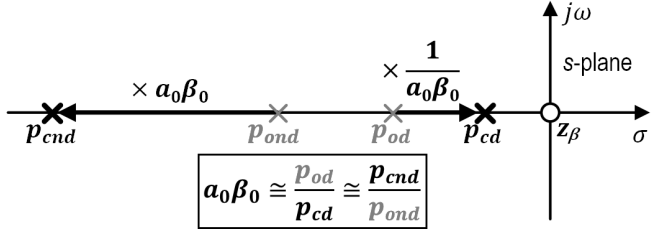}
\caption{Pole-splitting relation}
\label{PoleSplitRelation}
\end{figure}

\textit{Proof:}
For a given condition, $a(s)\beta(s)$ can be written 
\begin{align}
 a(s)\beta(s) = \frac{a_0\beta_0\frac{s}{\abs{p_{od}}}}{\left(1+\frac{s}{\abs{p_{od}}}\right)\left(1+\frac{s}{\abs{p_{ond}}}\right)}. 
\end{align}
The closed-loop poles of the circuit [or equivalently the zeros of the characteristic equation $1 + a(s)\beta(s)$] can be found as the zeros of 
\begin{align}
P(s)  = 1+s\frac{a_0\beta_0}{\abs{p_{od}}} +s^2\frac{1}{\abs{p_{od}p_{ond}}}.  \label{LT3} 
\end{align}
The two zeros of $P(s)$, $p_{cd}$ and $p_{cnd}$, are real and widely spaced (i.e., $\abs{p_{cnd}} \gg \abs{p_{cd}}$). This can be verified by comparing~(\ref{LT3}) with the standard form of the second-order polynomial with a damping ratio $\xi$ and a natural frequency $\omega_n$ given by 
\begin{align}
S(s)  = 1+s\frac{2\xi}{\omega_n} +s^2\frac{1}{\omega_n^2}.  \label{Standard} 
\end{align}
Equating (\ref{LT3}) with (\ref{Standard}) yields 
\begin{align}
\xi  = \frac{1}{2}\sqrt{\frac{p_{ond}}{p_{od}}} a_0\beta_0  >  \frac{1}{2}a_0\beta_0  \gg 1.
\end{align}
Thus, we can write
\begin{align}
P(s) \simeq 1+s\frac{1}{\abs{p_{cd}}} +s^2\frac{1}{\abs{p_{cd}p_{cnd}}}.  \label{LT4} 
\end{align}
Equating the coefficient of $s$ in~(\ref{LT3}) and (\ref{LT4}) results in 
\begin{align}
\abs{p_{cd}} \simeq \frac{\abs{p_{od}}}{a_0\beta_0}.  \label{pd} 
\end{align}
Similarly, by equating coefficients of $s^2$ in~(\ref{LT3}) and (\ref{LT4}) and using~(\ref{pd}), $p_{cnd}$ can be estimated as
\begin{align}
\abs{p_{cnd}} \simeq  \abs{p_{ond}} a_0\beta_0.  \label{pnd} 
\end{align}
Using (\ref{pd}), (\ref{pnd}) and eliminating the absolute functions, the pole-splitting relation given by (\ref{PoleSplit}) is established and is illustrated in Fig.~\ref{PoleSplitRelation}.    
\begin{flushright}
$\square$
\end{flushright}
 
\section{}
Applying the quadractic formula to the quadratic function in the denominator of~(\ref{ANMC}), $p_{cnd1}$ and $p_{cnd2}$ are 
\begin{align}
p_{cnd1}, p_{cnd2}  = -\frac{g_{m2}}{2C_{2}} \pm \frac{g_{m2}}{2C_{2}} \sqrt{1 - 4\frac{g_{m1}}{C_{c1}} \frac{C_2}{g_{m2}}}.  \label{formula}  
\end{align}
Using~(\ref{po2_nmc}) and~(\ref{GBW1}), we can write
\begin{align}
p_{cnd1}, p_{cnd2}  \simeq \frac{p_{o2}}{2} \mp \frac{p_{o2}}{2} \sqrt{1 + \frac{8 \textrm{GBW}}{p_{o2}} }.  \label{formula}  
\end{align}
Thus, when $p_{o2} = -8$GBW, $p_{cnd1}, p_{cnd2} \simeq p_{o2}/2 = -4 $GBW. And, when $p_{o2} = -4$GBW, $p_{cnd1}, p_{cnd2} \simeq p_{o2}/2 \mp jp_{o2}/2      = -2$GBW $\mp j2$GBW.

\bibliographystyle{IEEEtran}
\bibliography{IEEEabrv,References}

\begin{IEEEbiography}[{\includegraphics[width=2in,height=1.25in,clip,keepaspectratio]{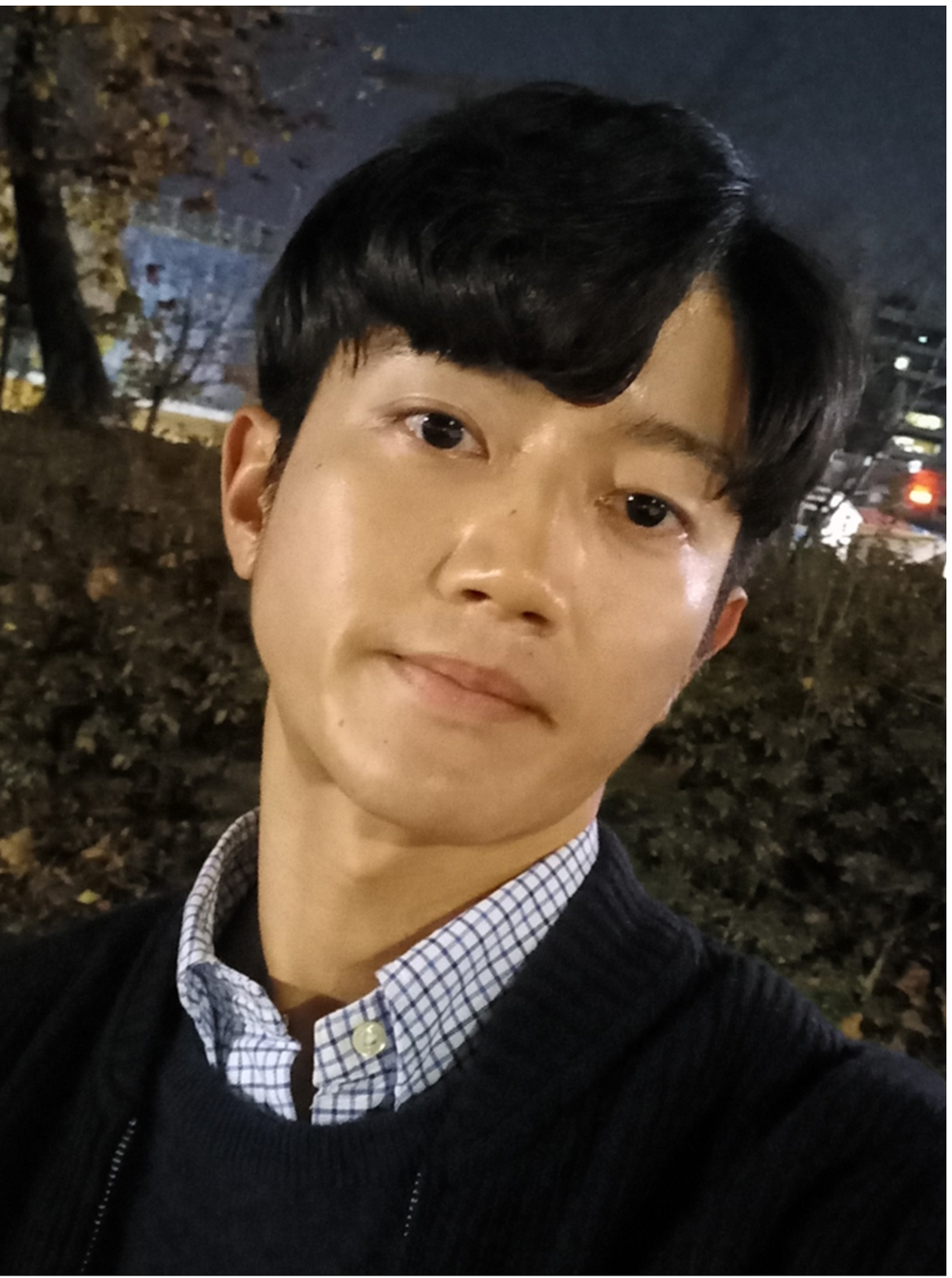}}]{Myungjun Kim} received the Ph.D. degree in electrical engineering from Korea Advanced Institute of Science and Technology (KAIST), Daejeon, Korea, in Feb, 2022.

His Ph.D. study has focused on the stability analysis and the development of the methodology for frequency compensation of various \textit{linear} analog integrated circuits (ICs) such as Badngap reference and (output-capacitorless) low-dropout regulator. His research interests also include the exploration for (Hopf) bifurcation and chaos that can occur in modern analog CMOS ICs from the \textit{nonlinear} dynamical system theory standpoint. 

He is currently working at the Memory Division, Samsung Electronics, Hwaseong, South Korea, as an analog IC designer. Dr. Kim also serves as a reviewer of \textsc{IEEE Transactions on Circuits and Systems—I: Regular Papers} and \textsc{IEEE Transactions on Circuits and Systems—II: Express Briefs}.

\end{IEEEbiography}

\end{document}